# *Charge Transport in One Dimension: Dissipative and Non-Dissipative Space-Charge Limited Currents*

## S R Holcombe* and E R Smith

**Mathematics Department, La Trobe University, Bundoora, Victoria, Australia 3083**

*Corresponding Author:
Email: srholcombe@gmail.com
Tel: +61 3 9479 2622

## *Abstract*

We consider charge transport in a pore where the dielectric constant inside the pore is much greater than in the surrounding material, so that the flux of the electric fields due to the charges is almost entirely confined to the pore. We develop exact solutions for the one component case for Dirichlet and Neumann boundary conditions using a Hopf-Cole transformation, Fourier series, and periodic solutions of the Burgers equation. These are compared with a simpler model in which the scaled diffusivity is zero so that all charge motion is driven by the electric field. In this non-dissipative case, recourse to an admissibility condition is used to obtain the physically relevant weak solution of a Riemann problem concerning the electric field. It is shown that the admissibility condition is Poynting's theorem.

# 1. Introduction

The development of nanopore systems in recent years has led to a considerable increase of interest in the dynamics of quasi-one dimensional systems. One interesting area of these studies has been charge transport in nanopores. For several systems of direct physical interest, charges move in a pore where the pore background has a high dielectric constant (i.e. that of water, or electrolyte with $\varepsilon \approx 80$) while the surround of the pore has a much lower dielectric constant (e.g. that of a lipid, with $\varepsilon \approx 2$), a ratio we will term as $\Delta$. As Kamenev et al [1] have pointed out, in these circumstances, the electric field due to the charges within the pore is almost all confined to the interior of the pore. That means that the electric field (or potential) due to the charges and any boundary conditions is a solution of a one dimensional Poisson equation. However the work of Barcilon, Chen and Eisenberg [2,3,4] show that induced charge on the pore walls can in fact have a non negligible impact upon stationary states within the pore, leading to so-called non-classical Poisson equations. One of the main results of their work was to show that the induced surface charge has a particularly strong impact for a specific relation of the dielectric ratios and the ratio of the pore radius to its length (aspect ratio). In what follows, we will assume a relatively inert and squat pore where higher order correction terms may be neglected and the assumption of complete field confinement reasonably maintained. We therefore from the outset exclude infinite or semi-infinite systems. Under these constraints it would appear that such a model, in reference to a space-charge limited diffusion/current within nanopores is perhaps most appropriate for reasonably inert and synthetic nanopores (Ramirez et al [5]), although there is still some debate as to such continuum models being appropriate for biological pores.

Another important observation of Kamenev et al [1] is that viscous damping of charged particle motion by the electrolyte background means that the motion of charges in a nanopore of the type considered will be diffusive, rather than governed by Newtonian dynamics. Barcilon et al and Ramirez et al, also present arguments for the validity of a diffusive description of carrier motion through the pore, and they in fact describe the motion of the carriers via the so-called Poisson-Nernst-Planck equations (PNP equations), a set of coupled nonlinear partial differential equations. In particular Barcilon et al examines steady states for various constant Dirichlet boundary conditions and various numbers of species. In previous work we have utilised the same set of equations (albeit in the non-dissipative form) to describe carrier motion over the surface of insulating materials. There we examined the dynamics of the carrier diffusion under various boundary conditions [6,7], extending the work of Shockley, Many, Rosen and others [8,9,10] to space-charge limited currents within insulating materials [11]. While the equations have direct applications [1] for charge distributions in nanopores, they more generally form a model for certain macroscopic descriptions of electrodiffusion, and a means for testing theories of mixed electrolytes as they diffuse. The equations may be used, for example, to examine theories of liquid junction potentials [12]. Here, as in our previous examination of surface carrier densities, we will be concerned predominantly on obtaining exact solutions for the time dependent carrier density of species within the system, focusing upon the mathematical requirements for determining unique solutions for the non-dissipative case.

To this end we shall consider initial charge distributions which are confined to a finite length pore (necessary to maintain the validity of the first order expansion) and either constant Dirichlet or Neumann boundary conditions on the electric potential at the ends of the pore. Immediately there is a wide difference between the situations in the two different boundary conditions. With Dirichlet boundary conditions, there is no significant constraint on the initial charge distribution. However, very general theory of partial differential equations (Garabedian [13]) shows that a necessary condition for the existence of a solution to the Poisson equation with Neumann boundary conditions is that the system must have net charge determined by the boundary conditions. In one dimension, this means that the difference between the values of electric field between the two ends of the system must be related to the net charge in the system.

With Dirichlet boundary conditions, it appears that the charge in the system spreads out to the ends of the pore and leaks from the system through the ends. With the potential difference along the pore zero, the long term state of the system appears to be zero charge density. On the other hand, with Neumann boundary conditions (specifying the electric field at either end), it turns out that some of the charge density has to remain confined to the pore. Thus the long term state of the system will retain some charge, as determined by the field difference from one end of the pore to the other. We will observe that the constant Neumann boundary conditions used here in fact describe a system in which there is no current over the ends of the domain (a closed pore or system). The one-dimensional Poisson and Nernst-Planck equations used with such boundary conditions may be found, for example, in [14] where the electrodiffusion of ionic species through concentric cylinders represents the extracisternal space of the outer hair cell. In terms of electrolytes this boundary condition is perhaps of more general interest for observing how the PNP system deals with diffusing species under the constraint of zero flux over the boundary, or over particular sections of the boundary [15]. For an example where both boundary conditions are used in the limiting case of zero diffusivity (i.e. non-dissipative case) see [6,7], which has direct physical application to measuring the resistivity of materials [11].

In this paper we begin by first obtaining the dynamic PNP equations. In section 2, we show how the potential along the pore with an external to internal dielectric constant ratio $\Delta \ll 1$ satisfies a one dimensional Poisson equation with corrections that are $O(\Delta)$ and then ignored in the rest of the paper. Our derivation somewhat follows that in [4] and we include it here for completeness. We then develop the model equations in section 3. There are, to begin, M ionic species present with various charges per ion, all an integer times an elementary charge Q. For each of them separately the charge current and charge density obey a simple conservation equation. The electric field obeys a one dimensional Poisson equation, which can be separated into M separate equations for the potential due to each species. Finally, it is assumed that the charge current due to each species is composed of two terms, one proportional to the charge density due to the species and the total electric field, the other term proportional to the concentration gradient for the species concerned times a diffusion constant. In section 3 we also scale the equations for the model so they have only two dimensionless parameters for each species, one a scaled diffusion constant and one a scaled version of the constant of proportionality in the charge-density-times-field term within the flux equations. The equations are developed as a set of coupled viscous Burgers' equations [16].

In the rest of the paper, we consider the simplest case where there is only one species present, highlighting the various subtleties present in obtaining solutions for the different types of boundary condition. In section 4 we present an exact solution for this simple one component system which uses a Hopf-Cole transformation [17,18] to convert the equation to a linear diffusion equation. We illustrate behaviour for some example initial states focusing on symmetric systems in order to simplify the analysis. We present numerical examples for a finite length system in section 5, using an initial charge density which is a delta function, and compare the diffusion for both boundary conditions to systems which diffuse linearly.

In section 6 we consider a one species system in which the diffusive terms are zero. This is a rather different model appropriate for diffusion of charge on surfaces of dielectrics (Mott and Gurney [19] and Holcombe et al [6,7,11]) and corresponds to an inviscid Burgers equation. The one species system can sustain discontinuities in the charge density or its gradient. We develop the solutions using the method of characteristics [20]. One interesting point about these systems is that they show that the scaled diffusivity in the equations of sections 3 and 4 behaves as a singular perturbation of the zero diffusivity case. Indeed we use this fact to eliminate certain ambiguities in the "weak" solutions of the inviscid system. We then appeal to the theory of hyperbolic equations to show that this method of vanishing viscosity may be replaced with a more fundamental observation concerning conservation of energy within the electric field. We in fact argue that Poynting's theorem is indeed the correct admissibility condition for determining weak solutions. The paper ends with a discussion in section 7.

## 2. The one dimensional Poisson equation

In this preliminary investigation we begin by considering the potential due to a charge q in a pore (axis along the X axis) of radius a, where we assume the induced charge on the surface of the pore to be negligible to the overall field within the pore. The background inside the pore has dielectric constant $\varepsilon_p$ while the material of the wall of the pore (assumed here to extend to infinity) has a dielectric constant $\varepsilon_w$. We consider pores in this paper for which the dimensionless ratio $\Delta = \varepsilon_w/\varepsilon_p$ obeys $0 \leq \Delta \ll 1$. The potential at $\mathbf{R}=(X,Y,Z)=(X,0,0)+\mathbf{r}$ with $\mathbf{r}=(0,Y,Z)$ due to such a charge at $\mathbf{R}_0=(X_0,Y_0,Z_0)$ obeys a Poisson equation

$$\nabla^2 \Phi_p(X,Y,Z;X_0,Y_0,Z_0;\Delta) = -\frac{4\pi q}{\varepsilon_p}\delta(X-X_0)\delta(Y-Y_0)\delta(Z-Z_0) \qquad (2.1)$$

inside the pore. In the wall of the pore the potential satisfies $\Phi_w$ a Laplace equation. Both of these potentials may be expanded in powers of $\Delta$ as

$$\Phi(\mathbf{R},\mathbf{R}_0;\Delta) = \sum_{n=0}^{\infty} \Delta^n \Phi_n(\mathbf{R},\mathbf{R}_0). \qquad (2.2)$$

So that $\Phi_{0,p}$ will be the first term of a convergent expansion for the potential in the pore, which will be accurate for the whole potential when $0 \leq \Delta \ll 1$. The boundary conditions on the potential are standard, thus at the internal surface we have, with $\mathbf{n}$ a unit normal vector to the surface,

$$\Phi_p(\mathbf{R},\mathbf{R}_0;\Delta) = \Phi_w(\mathbf{R},\mathbf{R}_0;\Delta) \quad \Rightarrow \quad \Phi_{n,p}(\mathbf{R},\mathbf{R}_0) = \Phi_{n,w}(\mathbf{R},\mathbf{R}_0) \text{ for } n \geq 0 \qquad (2.3)$$

and

$$\begin{aligned}\varepsilon_p \mathbf{n}\cdot\nabla\Phi_p(\mathbf{R},\mathbf{R}_0;\Delta) &= \varepsilon_w \mathbf{n}\cdot\nabla\Phi_w(\mathbf{R},\mathbf{R}_0;\Delta) \\ \Rightarrow \quad \mathbf{n}\cdot\nabla\Phi_{n,p}(\mathbf{R},\mathbf{R}_0) &= \mathbf{n}\cdot\nabla\Phi_{n-1,w}(\mathbf{R},\mathbf{R}_0) \quad \text{for } n \geq 1\end{aligned} \qquad (2.4)$$

while for n = 0, we also have

$$\mathbf{n}\cdot\nabla\Phi_{0,p}(\mathbf{R},\mathbf{R}_0) = 0. \qquad (2.5)$$

The specification of the potential is completed by requiring that $\Phi_w \to 0$ as $|\mathbf{r}| \to \infty$. Next we introduce a potential averaged across the pore as

$$\bar{\Phi}(X,X_0) = \frac{1}{A^2}\int_C d^2\mathbf{r}_0 \int_C d^2\mathbf{r}\, \Phi_{0,p}(\mathbf{r},X,\mathbf{r}_0,X_0) \qquad (2.6)$$

where C is the cross section of the pore and A the area of this cross section. We then have

$$\begin{aligned}\frac{\partial^2}{\partial X^2}\bar{\Phi}(X,X_0) &= \frac{1}{A^2}\int_C d^2\mathbf{r}_0 \int_C d^2\mathbf{r}\, \frac{\partial^2}{\partial X^2}\Phi_{0,p}(\mathbf{r},X,\mathbf{r}_0,X_0) \\ &= \frac{1}{A^2}\int_C d^2\mathbf{r}_0 \int_C d^2\mathbf{r}\left\{-\frac{4\pi q}{\varepsilon_p}\delta(\mathbf{r}-\mathbf{r}_0)\delta(X-X_0) - \nabla_\mathbf{r}^2 \Phi_{0,p}(\mathbf{r},X,\mathbf{r}_0,X_0)\right\}.\end{aligned} \qquad (2.7)$$

The integral over $\mathbf{r}$ in the second form of equation (2.7) has two terms. The second, with the Laplacian of $\Phi_{0,p}$ is zero by Gauss's theorem and the boundary condition on the gradient of the potential $\Phi_{0,p}$ if $x \neq x_0$ (for then $\Phi_{0,p}$ is sufficiently differentiable for Gauss's theorem to apply). On the other hand, if $x = x_0$, we may separate the integral on **r** over C\ $D_\delta$ where $D_\delta$ is a circle of radius $\delta$ with centre at $\mathbf{r}_0$ plus an integral over $D_\delta$. Using Gauss's theorem the integral over C\ $D_\delta$ is equal to a surface integral over the surface of $D_\delta$. We may separate the potential $\Phi_{0,p}$ into a term proportional to $1/|\mathbf{r} - \mathbf{r}_0|$ plus a sufficiently differentiable term. This last term has an integral over $D_\delta$ which exactly cancels its (inwards) surface integral over the surface of $D_\delta$. We may then calculate explicitly the contributions of the singular term over $D_\delta$ and the inwards surface integral over the surface of $D_\delta$. They give zero. Thus this second term in equation (2.7) is zero. The first term in equation (2.7) is simply evaluated to give

$$\frac{\partial^2}{\partial X^2} \bar{\Phi}(X, X_0) = -\frac{4\pi}{\varepsilon_p} \frac{q}{A} \delta(X - X_0). \tag{2.8}$$

Thus this average potential does indeed satisfy a one dimensional Poisson equation. We shall use this potential in the rest of the paper. The gross mismatch of dielectric constants between the interior of the pore and the wall does ensure that the electric flux due to the charges inside the pore is, to leading order in $\Delta$, confined to the pore as pointed out by Kamenev et al [1]. For a system with one dimensional density of one dimensional charge $\sigma(X,t)$, the mean potential will satisfy

$$\frac{\partial^2}{\partial X^2} \bar{\Phi}(X, t) = -\frac{4\pi}{\varepsilon_p} \sigma(X, t). \tag{2.9}$$

## 3. The system and its equations

### 3a. The equations:

We consider a system of M species of charges $Qq_k$ with the $q_k$ being integers for $1 \leq k \leq M$ with particle densities on the line $-L \leq X \leq L$ which are $\rho_k(X,t)$. We may often refer to these particle densities as carrier densities, although the terms "charge density" and "carrier density" will almost always be interchangeable. These species give rise to M current densities $J_k(X,t)$. There is a Poisson equation

$$\frac{\partial}{\partial X} E(X, t) = \frac{4\pi Q}{\varepsilon_p} \sum_{k=1}^{M} q_k \rho_k(X, t) \tag{3.1}$$

with some boundary conditions on it, either Neumann or Dirichlet, at $X = \pm L$. Ignoring boundary conditions for the moment, we introduce M separate Poisson equations

$$\frac{\partial}{\partial X} E_k(X, t) = \frac{4\pi Q}{\varepsilon_p} q_k \rho_k(X, t) \qquad \text{for } 1 \leq k \leq M. \tag{3.2}$$

Particle current equations of continuity are symmetric, but for the charge currents we have

$$\frac{\partial}{\partial X} J_k(X, t) + Qq_k \frac{\partial}{\partial t} \rho_k(X, t) = 0 \qquad \text{for } 1 \leq k \leq M. \tag{3.3}$$

Finally we have the Nernst-Planck equations

$$J_k(X,t) = -\Gamma_k Q q_k \frac{\partial}{\partial X}\rho_k(X,t) + \mu_k Q \rho_k(X,t)\sum_{L=1}^{M} E_L(X,t) \quad \text{for } 1 \le k \le M, \tag{3.4}$$

where $\Gamma_k$ are diffusion coefficients and $\mu_k$ are what we will term conductivity coefficients based on the fact that Ohms Law is $J = \sigma E$ where $\sigma$ is the conductivity. In certain circumstances $\sigma$ may be related to a thermodynamic diffusion coefficient via the Einstein relation or indeed, and perhaps more appropriately, the mobility of the species concerned.

### 3b. Scaling the equations:

First we define $X = Lx$ and $t = t_0\tau$ so that the domain of the problem is $-1 \le x \le 1$ and $\tau \ge 0$. We scale the particle densities, electric fields and current by, for $1 \le k \le M$,

$$\rho_k(X,t) = \frac{N_0}{2L}\nu_k(x,\tau), \quad E_k(X,t) = \frac{2\pi Q N_0}{\varepsilon_p}e_k(x,\tau) \quad \text{and} \quad J_k(X,t) = \frac{Q^2 N_0^2 \pi \mu_0}{\varepsilon_p L}j_k(x,\tau). \tag{3.5}$$

We also define

$$\mu_0 = \frac{1}{M}\sum_{k=1}^{M}\mu_k \quad \text{and} \quad \mu_k = \mu_0(1+\lambda_k) \Rightarrow \lambda_k = \frac{\mu_k - \mu_0}{\mu_0}, \quad \gamma_k = \Gamma_k\frac{QN_0}{2L^2 J_0} \quad \text{for } 1 \le k \le M \tag{3.6}$$

and note that the $\lambda_k$ are dimensionless. We have then

$$\frac{\partial}{\partial x}e_k(x,\tau) = q_k \nu_k(x,\tau) \quad \text{and} \quad e(x,\tau) = \sum_{k=1}^{M}e_k(x,\tau), \tag{3.7}$$

$$\frac{\partial}{\partial x}j_k(x,\tau) + q_k \frac{\partial}{\partial \tau}\nu_k(x,\tau) = 0 \tag{3.8}$$

and

$$j_k(x,t) = -\gamma_k q_k \frac{\partial}{\partial x}\nu_k(x,\tau) + (1+\lambda_k)\nu_k(x,\tau)\sum_{L=1}^{M}e_L(x,\tau). \tag{3.9}$$

The dimensionless parameters are then scaled differences in current flux rates, $\lambda_k$, and the M scaled diffusion constants $\gamma_k$. These scaled diffusion constants are arbitrary real numbers with $\gamma_k \ge 0$, while the $\lambda_k$ are arbitrary. Note that each scaled field $e_k$ may be written as the x derivative of a scaled potential $\varphi_k$.

### 3c. Partial reduction of the equations:

If we substitute equation (3.9) into (3.8), and use equation (3.7), we obtain

$$\frac{\partial}{\partial x}\left\{\frac{\partial}{\partial \tau}e_k(x,\tau) - \gamma_k \frac{\partial^2}{\partial x^2}e_k(x,\tau) + \frac{(1+\lambda_k)}{q_k}\left(\frac{\partial}{\partial x}e_k(x,\tau)\sum_{L=1}^{M}e_L(x,\tau)\right)\right\} = 0. \tag{3.10}$$

Now each electric field is minus the gradient of a potential, $e_k(x,\tau) = -\partial \varphi_k(x,\tau)/\partial x$ and

$$\phi_k(x,\tau) = \phi_k(-1,\tau) + \int_{-1}^{x} e_k(y,\tau)dy, \qquad (3.11)$$

so that we also have, for 1 ≤ k ≤ M

$$\frac{\partial}{\partial x}\left\{\frac{\partial}{\partial x \partial \tau}\phi_k(x,\tau) - \gamma_k \frac{\partial^3}{\partial x^3}\phi_k(x,\tau) - \frac{(1+\lambda_k)}{q_k}\left(\frac{\partial^2}{\partial x^2}\phi_k(x,\tau)\sum_{L=1}^{M}\frac{\partial}{\partial x}\phi_L(x,\tau)\right)\right\} = 0. \qquad (3.12)$$

These M nonlinear equations are not easy to solve. We can however solve them for M=1, which we shall do in section 4. This method relies crucially on being able to write the left hand side of equation (3.12) as a second derivative with respect to x. One procedure that is fruitful for the one component case is to re-write equation (3.12) as, for 1 ≤ k ≤ M,

$$\frac{\partial^2}{\partial x^2}\left\{\frac{\partial}{\partial \tau}\phi_k(x,\tau) - \gamma_k \frac{\partial^2}{\partial x^2}\phi_k(x,\tau) - \frac{(1+\lambda_k)}{2q_k}\left(\frac{\partial}{\partial x}\phi_k(x,\tau)\right)^2\right\}$$
$$= \frac{(1+\lambda_k)}{q_k}\frac{\partial}{\partial x}\left(\frac{\partial^2}{\partial x^2}\phi_k(x,\tau)\sum_{\substack{L=1\\L\neq k}}^{M}\frac{\partial}{\partial x}\phi_L(x,\tau)\right). \qquad (3.13)$$

The left hand side of this equation is the second x derivative of an object nonlinear in $\varphi_k$ and its derivatives. The exact one component solutions that we develop in section 4 rely on being able to write the whole equation as a second x derivative being equal to zero. It is not immediately obvious how to write the right hand side of equation (3.13) as a second x derivative. It is reasonably simple to make linear combinations of the $\varphi_k$ (the linear combination is, in fact, the sum of all the $\varphi_k$) which give at least one equation whose left hand side may be written as a second derivative with respect to x of an object, but not apparently possible to find M such linear combinations.

One solution to this difficulty is to use numerical solutions of the system. This and further analysis of multi-component systems is in preparation. It is useful to note that the dimensionless reduction of the equations for the system makes the range of parameters for numerical investigation of "typical" systems much smaller.

### *3d. The electric fields:*

There are two possible simple boundary conditions on the electric field and its potential, Dirichlet boundary conditions and Neumann boundary conditions. In Dirichlet boundary conditions we have $\varphi(\pm 1,\tau)$ specified where $e = -\partial\varphi/\partial x$ and in Neumann boundary conditions we have $e(\pm 1,\tau)$ specified with constraints on the values depending on the net charge on the system. It is interesting to note that while the Dirichlet problem on a finite domain can have any values for $\varphi(\pm 1,\tau)$, there is also a constraint on the values of $\varphi(\pm\infty,\tau)$ if we have a problem on a doubly infinite domain.

#### *Boundary conditions (i): the Neumann case*

In this case, if the initial total charge on the system is $Q_0$, then we have

$$e(1,\tau) = e(-1,\tau) + \sum_{k=1}^{M} q_k \int_{-1}^{1} v_k(x,\tau)dx = e(-1,\tau) + Q_0(\tau) \qquad (3.14)$$

so that the total charge in the system determines the difference between the electric fields at x = ±1. The potential in the system is then

$$\phi(x,\tau) = -e(-1,\tau)(x+1) - \sum_{k=1}^{M} q_k \int_{-1}^{x} dy \int_{-1}^{y} v_k(z,\tau) dz \qquad (3.15)$$

where we have made the (quite general) assumption that $\varphi(-1,\tau)=0$. Notice that if we define $e(-1,\tau)$ and $e(+1,\tau)$ to be fixed in time, then not only is the initial total charge on the system defined by the difference between the electric fields at either end of the system, this total charge must be conserved so that charge cannot leak out of the system. This suggests that at large time the charge distributions will reach a steady state. The steady state is not generally one where each of the densities $v_k(x,\infty)$ is constant. To see this note that if the densities reach a steady state, then from (3.8), we have $\partial j_k(x,\tau)/\partial x = 0$ and so each current density is therefore constant in x. Since charge can not leak from the system, $j_k(\pm 1,\tau) = 0$ and hence $j_k(x, \infty) = 0$. Equation (3.9) then gives

$$\gamma_k q_k \frac{\partial}{\partial x} v_k(x,\infty) = (1+\lambda_k) v_k(x,\infty) \sum_{L=1}^{M} e_L(x,\infty). \qquad (3.16)$$

Assume now that each carrier density in the steady state is of constant distribution in x. Under such assumptions

$$(1+\lambda_k) v_k(x,\infty) e(x,\infty) = 0 \qquad (3.17)$$

for each k. Since $e(x,\infty)$ can not be zero for all x, otherwise we would have no charge in the system, we must have then $(1+\lambda_k)v_k(x,\infty) = 0$ for each k and all x. This is in contradiction of the initial assumption that the system contains charge.

### *Boundary conditions (ii): the Dirichlet case*

The potential in a system is only defined up to an additive constant, so we assume $\varphi(-1,\tau) = \varphi_k(-1,\tau) = 0$ for $1 \le k \le M$. We then have equation (3.15) again, but this time do not have values for $e(-1,\tau)$, yet. If we put x = 1 in equation (3.15), and integrate by parts, we obtain

$$\phi(1,\tau) = -2e(-1,\tau) - Q_0(\tau) + \mathbf{m}(\tau), \qquad (3.18)$$

where $Q_0(\tau)$ is the total charge in the system at time $\tau$ and $\mathbf{m}(\tau)$ is the dipole moment of the system at time $\tau$, namely

$$\mathbf{m}(\tau) = \int_{-1}^{1} y \, dy \sum_{k=1}^{M} q_k v_k(y,\tau). \qquad (3.19)$$

This then gives an equation for $e(-1,\tau)$ when $\varphi(1,\tau)$ is defined by boundary conditions. These equations will normally always have solutions, though there are obvious extra constraints when the interval for the system is the whole real line.

With Dirichlet conditions it is possible for charge to leak out of the system. We may calculate the amount of charge which has leaked from the system in scaled time $\tau$ by integrating $j_k(\pm 1,\tau)$ (equation (3.9)) from 0 to $\tau$. It is then simple to show that the total charge remaining in the system at time $\tau$ plus the amount of charge that has leaked from the system is equal to the amount of charge in the system at $\tau = 0$. The large time behaviour of the system then corresponds to an empty system with linear potential and constant field. These are indeed solutions of equation (3.13) for the potentials.

## 4. Exact solution for one component

If we have only one component present then we have $\lambda_1 = 0$ (by equation (3.6)) and we may write $\gamma_1 = \gamma$ so that equation (3.13) becomes (suppressing the label subscript and using $q_1=1$, allowed by the scaling process),

$$\frac{\partial^2}{\partial x^2}\left\{\frac{\partial}{\partial \tau}\phi(x,\tau) - \gamma\frac{\partial^2}{\partial x^2}\phi(x,\tau) - \frac{1}{2}\left(\frac{\partial}{\partial x}\phi(x,\tau)\right)^2\right\} = 0. \tag{4.1}$$

If we integrate twice with respect to x we obtain

$$\frac{\partial}{\partial \tau}\phi(x,\tau) - \gamma\frac{\partial^2}{\partial x^2}\phi(x,\tau) - \frac{1}{2}\left(\frac{\partial}{\partial x}\phi(x,\tau)\right)^2 = xf''(\tau) + g'(\tau) \tag{4.2}$$

where

$$f''(\tau) = \gamma\frac{\partial^2}{\partial x^2}e(-1,\tau) - \frac{\partial}{\partial \tau}e(-1,\tau) - e(-1,\tau)v(-1,\tau) \tag{4.3}$$

and

$$g'(\tau) = \frac{\partial}{\partial \tau}\phi(-1,\tau) - \gamma\frac{\partial^2}{\partial x^2}\phi(-1,\tau) - \frac{1}{2}\{e(-1,\tau)\}^2 + f''(\tau). \tag{4.4}$$

The two functions f and g in these three equations are written as derivatives because later we shall need their indefinite integrals to construct integrating factors. The actual functions are then defined by integrating on $[0,\tau]$ giving

$$f'(\tau) = \int_0^\tau f''(\tau')d\tau', \ f(\tau) = \int_0^\tau f'(\tau')d\tau' \text{ and } g(\tau) = \int_0^\tau g'(\tau')d\tau' \tag{4.5}$$

which imply $df(0)/dt = f(0) = g(0) = 0$. We also introduce

$$F(\tau) = \frac{1}{2}\int_0^\tau f'(\tau')^2 d\tau' \ \Rightarrow \ F(0) = 0. \tag{4.6}$$

Equation (4.2) is an inhomogeneous viscous Burgers equation. Burgers [16] introduced these equations to provide a model equation system where the development of turbulence might be observed, but later found reasons not to develop them as models of turbulence [21]. In later work he considered equations like (4.2) simply as nonlinear diffusion equations. He was much concerned in that work to develop solutions at what is here considered the diffusivity $\gamma$ (although commonly referred to as viscosity) in the limit of very small positive $\gamma$. In particular, he found that in the limit $\gamma = 0$, the system could sustain shock waves in the charge density. We shall see examples of this in section 6.

We now introduce a fairly standard Hopf-Cole transformation of equation (4.2). We define

$$\psi(x,\tau) = \exp\{(\phi(x,\tau) - g(\tau) - xf'(\tau) - F(\tau))/2\gamma\}. \tag{4.7}$$

If we multiply equation (4.2) by $\Psi/2\gamma$ and use this transform, we obtain

$$\frac{\partial}{\partial t}\psi(x,\tau)-\gamma\frac{\partial^2}{\partial x^2}\psi(x,\tau)-f'(\tau)\frac{\partial}{\partial x}\psi(x,\tau)=0 \tag{4.8}$$

The point of this Hopf-Cole transformation is now clear, it reduces the non linear equation for $\varphi$ into a linear equation for $\Psi$. The function $\Psi(x,\tau)$ has a Fourier series on $-1 < x < 1$ using a normalized basis of functions $A_n\exp(i\omega_n x)$. The particular choice will depend on the boundary conditions on $\Psi(x,\tau)$ at $x = \pm 1$ and it is then simple to show that this series is

$$\psi(x,\tau)=\sum_{n=-\infty}^{\infty}a_n\exp\left(-\omega_n^2\gamma\tau+i\omega_n\left(x+f(\tau)\right)\right) \tag{4.9}$$

where

$$\psi(x,0)=\sum_{n=-\infty}^{\infty}a_n\exp(i\omega_n x)=\exp(\phi(x,0)/2\gamma). \tag{4.10}$$

From equation (4.7), we obtain

$$e(x,\tau)=-f'(\tau)-2\gamma\frac{\partial\psi(x,\tau)/\partial x}{\psi(x,\tau)}, \tag{4.11}$$

and then

$$v(x,t)=2\gamma\left\{\left(\frac{\partial\psi(x,\tau)/\partial x}{\psi(x,\tau)}\right)^2-\frac{\partial^2\psi(x,\tau)/\partial x^2}{\psi(x,\tau)}\right\}. \tag{4.12}$$

It is also of interest to solve equation (4.2) on the infinite interval $(-\infty,\infty)$, because, as we shall see, its study is rather simpler and shows us part of what to expect of the solutions. Equation (4.3) is the same except that $-\infty$ replaces $-1$ in the arguments of the field and its derivatives. Since the amount of charge that can diffuse out to $-\infty$ in finite time is negligible, we have $\partial^2 f(\tau)/\partial\tau^2 = 0$ and so $f(\tau) = 0$ and $F(\tau) = 0$. From equation (4.4) we have

$$g'(\tau)=\frac{\partial}{\partial\tau}\phi(-\infty,\tau)-\gamma\frac{\partial^2}{\partial x^2}\phi(-\infty,\tau)-\frac{1}{2}\{e(-\infty,\tau)\}^2. \tag{4.13}$$

This may not necessarily be zero. With (cf. equation (4.7))

$$\psi(x,\tau)=\exp\{(\phi(x,\tau)-g(\tau))/2\gamma\} \tag{4.14}$$

we obtain (cf. equation (4.8)) the diffusion equation

$$\frac{\partial}{\partial t}\psi(x,\tau)-\gamma\frac{\partial^2}{\partial x^2}\psi(x,\tau)=0. \tag{4.15}$$

This has the solution

$$\psi(x,\tau)=\frac{1}{2\sqrt{\pi\gamma\tau}}\int_{-\infty}^{\infty}\psi(y,0)\exp\left(-\frac{(y-x)^2}{4\gamma\tau}\right)dy. \tag{4.16}$$

The field e and the charge density are then obtained from equations (4.11) and (4.12).

## 5. Simple examples

The first example we examine is on (-∞,∞) with an initial condition

$$\nu(x,0) = \alpha\delta(x) \tag{5.1}$$

and Neumann boundary conditions chosen so that the field at the left hand end is minus the field at the right hand end. While we are in fact violating a key assumption used in formulating the model, that being its applicability to finite systems, this is a reasonably simple problem which will give us qualitative information about the way the charge diffuses in symmetrical circumstances. It will also yield results that are mathematically applicable when it comes to analysing the non-dissipative system in section 6. The field from the initial carrier distribution is then

$$e(x,0) = \begin{cases} -\alpha/2 & \text{for } x < 0 \\ \alpha/2 & \text{for } x > 0 \end{cases}. \tag{5.2}$$

The initial potential is then

$$\phi(x,0) = -\frac{\alpha}{2}|x|. \tag{5.3}$$

This explains why we use Neumann boundary conditions. There are serious convergence problems if we try to use Dirichlet conditions on the infinite interval. From equation (5.3), we then obtain from equation (4.16),

$$\psi(x,\tau) = \frac{1}{2}e^{\alpha\gamma\tau/4}\left\{\exp\left(-\frac{\alpha x}{2}\right)\text{erfc}\left(-\frac{x-\alpha\gamma\tau}{2\sqrt{\gamma\tau}}\right) + \exp\left(\frac{\alpha x}{2}\right)\text{erfc}\left(\frac{x+\alpha\gamma\tau}{2\sqrt{\gamma\tau}}\right)\right\}. \tag{5.4}$$

If we use the asymptotic expansion erfc(x) ≈ exp(-x²)/(x√π), then we obtain the large positive x, finite τ asymptotic representation

$$\nu(x,\tau) = \frac{\alpha\gamma}{\sqrt{\pi\gamma\tau}}\exp\left(-\frac{(x-\alpha\gamma\tau)^2}{4\gamma\tau}\right)\left\{1 + O\left(\frac{1}{x}\right)\right\}. \tag{5.5}$$

Notice that the charge density decays to zero as x becomes large, but is finite for all large positive x. This reflects the capacity of diffusive processes to shift matter (charge) infinite distances in finite time, albeit only a very small amount of charge. This fact has implications for finite length systems because of the differences between the effects of the two simple boundary conditions on the potential. Thus, in the finite length system, charge may be expected to reach the ends of the interval at once and then begin leaking from the system, so long as the boundary conditions on the potential do not prevent charge leakage. If there are Neumann conditions on the potential for the finite length system, then as in this case, charge will immediately reach the ends of the system and then either begin diffusing back or accumulate, eventually reaching a steady state.

To illustrate these ideas we now consider systems with constant symmetric Neumann or Dirichlet boundary conditions, (for the Neumann case, $e(-1,\tau) = -e_0$ and $e(1,\tau) = e_0$, for the Dirichlet case $\varphi(\pm 1,\tau) = 0$) and a symmetric initial condition $v(x,0)=v(-x,0)$ (and we shall later consider the very simple case $v(x,0) = \alpha\delta(x)$). In this circumstance, we expect that the charge density should be an even function of x and the field should be an odd function of x. For a one component system, equation (3.10) is

$$\frac{\partial}{\partial x}\left\{\frac{\partial e}{\partial \tau} - \gamma\frac{\partial^2 e}{\partial x^2} + e\frac{\partial e}{\partial x}\right\} = 0. \tag{5.6}$$

If we integrate this from -1 to x we will obtain $\partial^2 f(\tau)/\partial\tau^2$ as defined in (4.3). Integrating then from x to 1 and noting that the electric field is odd, implies $\partial^2 f(\tau)/\partial\tau^2 = 0$, This makes development of the solution rather more simple. Thus, from equations (4.6),(4.7),(4.8) we have

$$\psi(x,\tau) = \exp\{(\phi(x,\tau) - g(\tau))/2\gamma\} \text{ and } \frac{\partial}{\partial t}\psi(x,\tau) - \gamma\frac{\partial^2}{\partial x^2}\psi(x,\tau) = 0, \tag{5.7}$$

and by equation (4.4), $g(\tau)$ may not be zero.

### 5a. Neumann Boundary Conditions:

For the case of Neumann boundary conditions, with

$$\alpha = \int_{-1}^{1} v(x,\tau)dx \tag{5.8}$$

we have $e_0 = \alpha/2$. Since the problem is symmetric in x, we need only solve (5.7) on $0 \le x \le 1$ with boundary conditions

$$\frac{\partial}{\partial x}\psi(0,\tau) = 0 \text{ and } \frac{\partial}{\partial x}\psi(1,\tau) + \beta\psi(1,\tau) = 0 \text{ where } \beta = \frac{\alpha}{4\gamma}. \tag{5.9}$$

We may then use the orthonormal basis set

$$h_n(x) = \sqrt{\frac{2(\beta^2 + \omega_n^2)}{\beta + \beta^2 + \omega_n^2}}\cos(\omega_n x) \tag{5.10}$$

where the $\omega_n$ are the positive solutions of

$$\omega_n \tan(\omega_n) = \beta. \tag{5.11}$$

For each integer $n \ge 0$, there is one solution $\omega_n$ in the interval $n\pi \le \omega_n < (n+1/2)\pi$.

$$\psi(x,\tau) = \sum_{n=-\infty}^{\infty} \psi_n \exp(-\omega_n^2\gamma\tau)\cos(\omega_n x) \tag{5.12}$$

where

$$\exp(\phi(x,0)/2\gamma) = \sum_{n=-\infty}^{\infty} \psi_n \cos(\omega_n x) \tag{5.13}$$

on 0 ≤ x ≤ 1. For large $\gamma\tau$, this series is dominated by the n = 0 term and we then obtain a steady state field and steady state density given by

$$\frac{2e}{\alpha} = \frac{\omega_0}{\beta}\tan(\omega_0 x) \text{ and } \frac{2\nu}{\alpha} = \frac{\omega_0}{\beta}\left(1+\tan^2(\omega_0 x)\right). \quad (5.14)$$

These steady state results are determined independent of the initial charge density, relying only on the total charge in the system at $\tau$ = 0. Further, they satisfy the steady state form of equation (5.6) when $f(\tau) = 0$. Figure 5.1 shows the steady state scaled density for five values of $\beta$. The reason we have chosen to present the field and density with factors $2/\alpha$ is that they allow us, using a rescaled time $T = \gamma\tau$, to compare the results with the linear problem. This corresponds to a linear diffusion problem for $\nu(x,T)$ with unit diffusion constant and $\partial\nu(1,T)/\partial x = 0$ (the second of these deriving from the requirement that the particle flux at x = 1 be zero). Thus the linear problem has solution

$$2\nu_L(x,T)/\alpha = \sum_{n=0}^{\infty} a_n \exp(-\pi^2 n^2 T)\cos(\pi n x) \quad (5.15)$$

with

$$a_n = \frac{2}{\alpha}\int_0^1 \nu_L(y,0)\cos(\pi n y)dy. \quad (5.16)$$

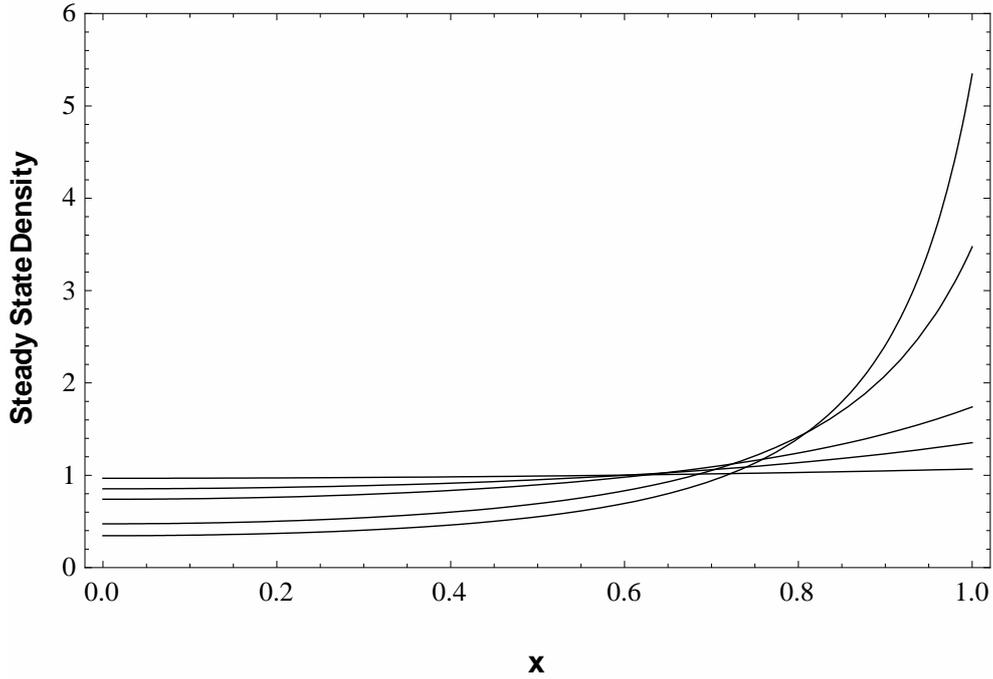

**Figure 5.1**: Scaled steady state charge density for $\beta$=0.1,0.5,1,3 and 5. The densities are increasing functions of x and the curves may be identified by the fact that at x=1, $2\nu/\alpha$ is an increasing function of $\beta$

We now consider the simple case where the initial carrier density is $\nu(x,0) = \alpha\delta(x)$, so that the net charge in the system is $\alpha$. We then have, for the nonlinear problem, $\varphi(x,0)/2\gamma = -\beta|x|$ since $g(0) = 0$. Thus

*13*

$$\frac{2}{\alpha}v(x,t) = \frac{\partial}{\partial x}\left\{\frac{\sum_{n=0}^{\infty}\frac{\omega_n}{\beta+\beta^2+\omega_n^2}\exp(-\omega_n^2\gamma\tau)\sin(\omega_n x)}{\sum_{n=0}^{\infty}\frac{\beta}{\beta+\beta^2+\omega_n^2}\exp(-\omega_n^2\gamma\tau)\cos(\omega_n x)}\right\}. \tag{5.17}$$

We also have

$$\frac{2}{\alpha}v_L(x,T) = 1 - 2\sum_{n=0}^{\infty}\exp(-\pi^2 n^2 T)\cos(\pi n x). \tag{5.18}$$

It is of some interest to see how the linear and nonlinear descriptions of the problem compare. As we shall see, they can be very different. For small T, the solutions can be very similar, while for large T the linear system settles to a steady state with constant density, while the nonlinear system settles to the steady state illustrated in Figure 5.1. These are illustrated in Figures 5.2 to 5.9. The linear model is given with the fine line, the nonlinear model with the heavy line.

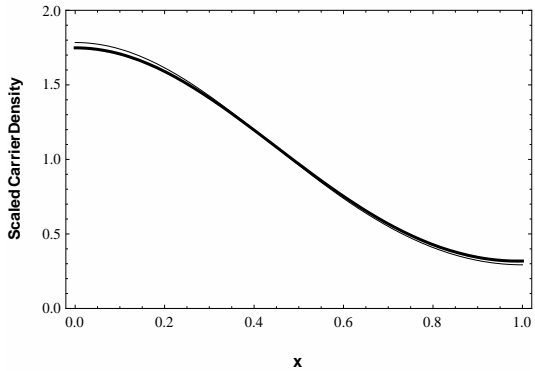

**Figure 5.2**: $\beta = 0.1$, T=0.1

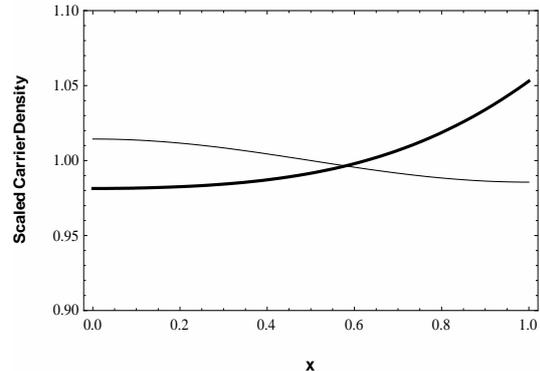

**Figure 5.3**: $\beta = 0.1$, T=0.5

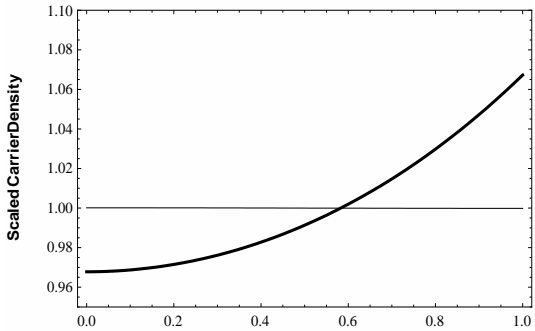

**Figure 5.4**: $\beta = 0.1$, T=0.9

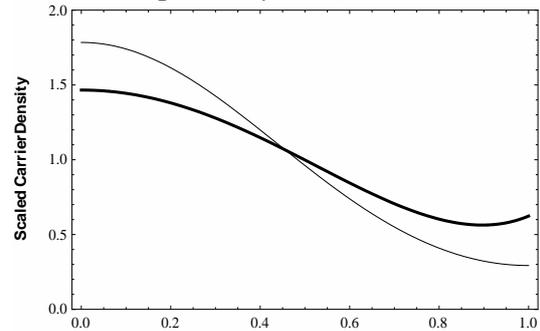

**Figure 5.5**: $\beta = 1$, T=0.1

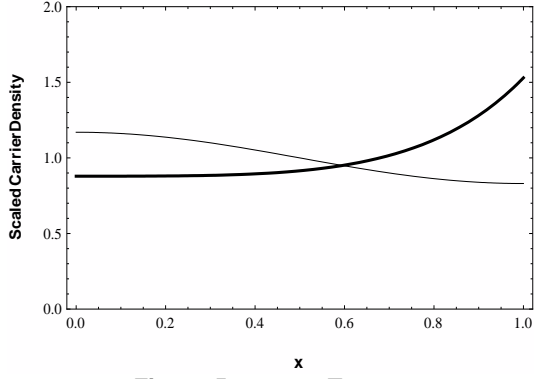
**Figure 5.6**: $\beta = 1$, T=0.25

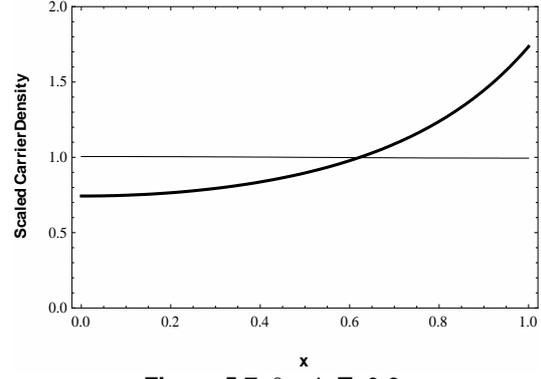
**Figure 5.7**: $\beta = 1$, T=0.6

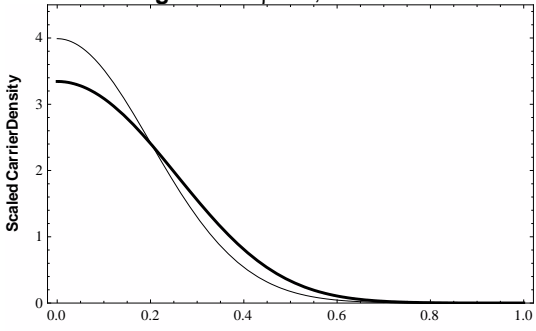
**Figure 5.8**: $\beta = 2$, T=0.02

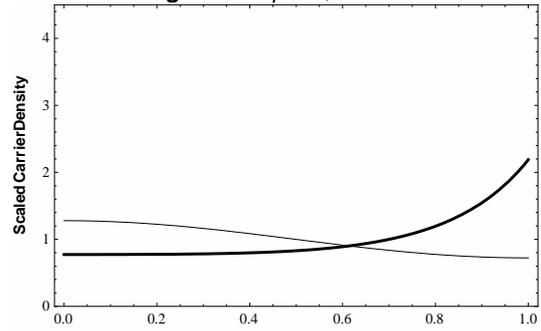
**Figure 5.9**: $\beta = 2$, T=0.2

It is clear that the system evolves rapidly to its steady state from a state very like the linear system. As $\beta$ increases the time required for the system to become close to the steady state decreases.

### 5b. Dirichlet Boundary Conditions:

For the Dirichlet case, we consider the same initial condition although specify the potential at the ends of the pore, insisting $\varphi(\pm 1,\tau) = 0$. We obtain therefore a set of conditions on x belonging to [0,1]

$$\psi(x,0) = \exp(\phi(x,0)/2\gamma), \quad \frac{\partial}{\partial x}\psi(0,\tau) = 0 \text{ and } \psi(1,\tau) = \exp(-g(\tau)/2\gamma), \quad (5.19)$$

where, upon taking into account $\partial^2 f(\tau)/\partial \tau^2 = 0$, the boundary conditions on the potential and the symmetry of the charge density,

$$g'(\tau) = \gamma v(1,\tau) - \frac{1}{2}e^2(1,\tau). \quad (5.20)$$

We cannot, however, specify the time dependence of the charge density or the electric field at $x=\pm 1$ as this is implicit to the problem. Rather we must perform a periodic extension of the initial charge density over the real line. Doing so allows us to utilise periodic solutions of the linear heat equation that in turn yield periodic solutions of Burgers' equation, ensuring $\varphi(\pm 1,\tau) = 0$. Such a solution representing the carrier evolution of the entire set of image charges is also representative of the evolution of the physical charge over x belonging to [-1,1].

As was done for the Neumann case we take an initial point charge density $\nu(x,0)=\alpha\delta(x)$. The initial potential associated with a configuration of image charges that ensures $\varphi(2n+1,\tau) = 0$ for $n=\pm 1,\pm 2,\ldots$ is therefore,

$$\phi_P(x_n,0) = \left(1 - |x_n - 4n|\right)\frac{\alpha}{2} \quad \text{for} \quad 2(n-1) < x_n < 2(n+1). \tag{5.21}$$

The triangular wave formation of this potential may be seen in Figure 5.10, where the delta-function charge density is represented by vertical "spikes".

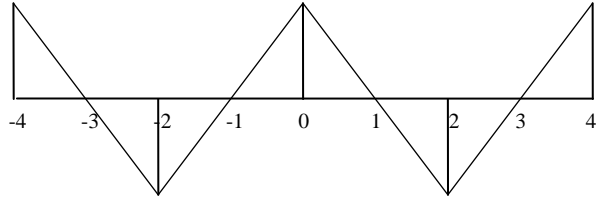

**Figure 5.10:** Periodic array of image charges (vertical spikes) establishing boundary conditions on the physical charge in x belonging to [-1,1]. The potential from the images is a triangular wave over the entire real line.

A solution of the heat equation on [-1,1] that is even about x=0 and that has the correct periodicity over the real line of images may be given by

$$\psi(x,\tau) = \frac{1}{2}a_0 + \sum_{k=1}^{\infty} a_k e^{-\frac{\gamma}{4}k^2\pi^2\tau}\cos\left(\tfrac{1}{2}k\pi x\right), \tag{5.22}$$

where

$$a_k = \int_0^2 \psi(x,0)\cos\left(\tfrac{1}{2}k\pi x\right).$$

The normalised or scaled carrier density over [0,1] is therefore

$$\frac{2}{\alpha}\nu(x,t) = \frac{1}{\beta}\frac{\partial}{\partial x}\left\{\frac{\frac{1}{2}\sum_{k=1}^{\infty}\dfrac{\left(e^{2\beta}+(-1)^{k+1}e^{-2\beta}\right)k\pi}{1+(k\pi/2\beta)^2}e^{-\frac{1}{4}k^2\pi^2 T}\sin\left(\tfrac{1}{2}k\pi x\right)}{\sinh(\beta)+\sum_{k=1}^{\infty}\dfrac{\left(e^{2\beta}+(-1)^{k+1}e^{-2\beta}\right)}{1+(k\pi/2\beta)^2}e^{-\frac{1}{4}k^2\pi^2 T}\cos\left(\tfrac{1}{2}k\pi x\right)}\right\}, \tag{5.23}$$

from which we may readily obtain

$$g(T) = -2\gamma\log\left(\frac{\sinh(\beta)}{\beta} + \frac{2\sinh(\beta)}{\beta}\sum_{k=1}^{\infty}\frac{(-1)^k e^{-k^2\pi^2 T}}{1+(k\pi/\beta)^2}\right). \tag{5.24}$$

It is not difficult to show

$$2\sum_{k=1}^{\infty}\frac{(-1)^k}{1+(k\pi/\beta)^2}=\frac{\beta}{\sinh(\beta)}-1,$$

and therefore verify g(0)=0. Now that we have g(T) it is a simple matter to show directly that the potential indeed satisfies (4.2) and the Dirichlet boundary conditions φ(±1,τ)=0.

As was previously done for the Neumann case we may compare the evolution of the nanopore carrier density against that of the linear system where the scaled linear carrier density may be shown to be

$$\frac{\alpha}{2}\nu_L(x,T)=\sum_{k=1}^{\infty}\left(1+(-1)^{k+1}\right)e^{-\frac{1}{4}k^2\pi^2 T}\cos\left(\tfrac{1}{2}k\pi x\right). \tag{5.25}$$

It is interesting to note that while both linear and nonlinear systems empty as T→ ∞, the carrier density for the nonlinear system at x=±1 is only zero for T=0 or when T→ ∞, whereas for the linear system it is zero for all T. This indicates a necessary asymmetry occurring between the positive and negative image charge densities in the nonlinear case. The asymmetry may be better illustrated by observing the image densities over time as shown in Figures 5.11 and 5.12.

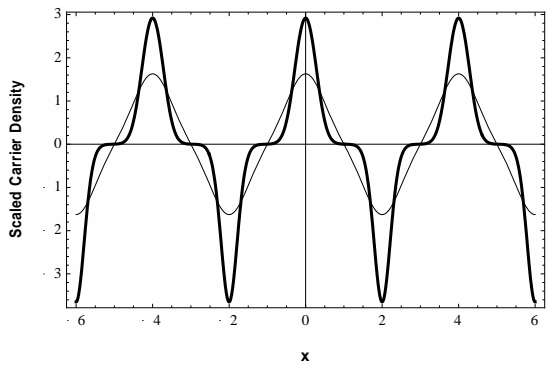
**Figure 5.11**: linear (fine) and nonlinear (heavy) image densities for T=0.03, β=1

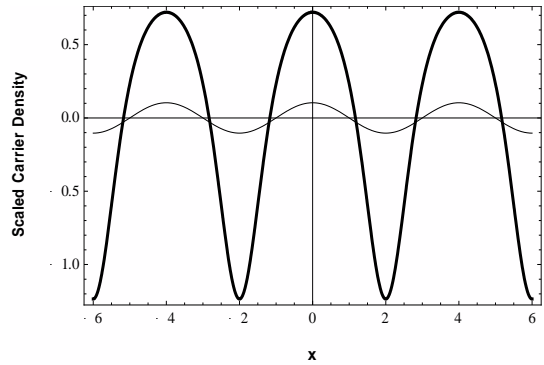
**Figure 5.12**: linear (fine) and nonlinear (heavy) image densities for T=0.3, β=1

As for the Neumann boundary conditions, we may compare the carrier evolution between the linear (fine line) and nonlinear system (heavy line) given in Figures 5.13 and 514.

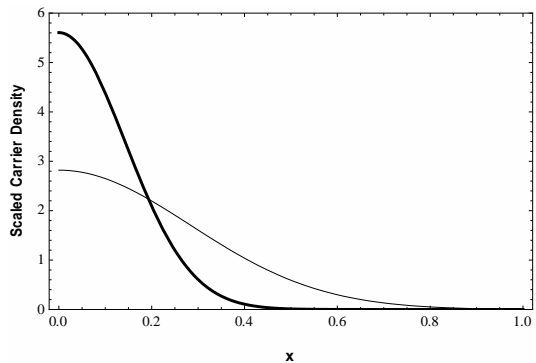
**Figure 5.13**: β=0.1,T=0.01

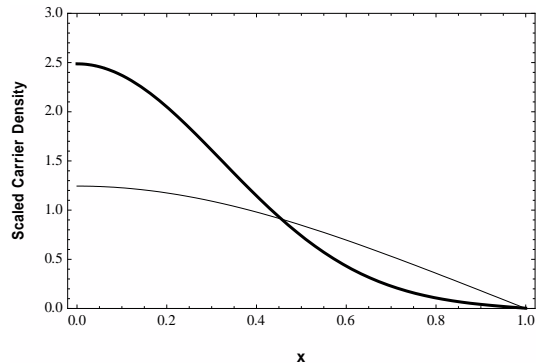
**Figure 5.14**: β=0.1,T=0.05

However in this case it is of more interest to observe the diffusion profiles for the carrier density over the pore and at the ends of the pore over time, and for various β as shown in Figures 5.15 and 5.16 where the profiles at x=1 increase with β.

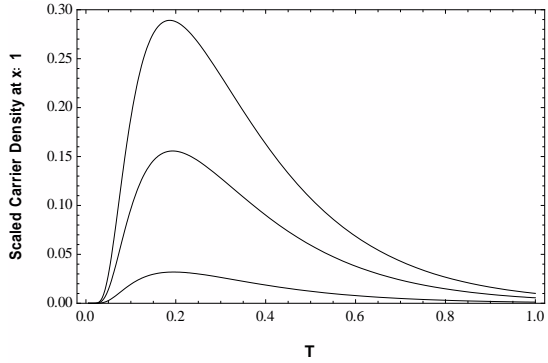 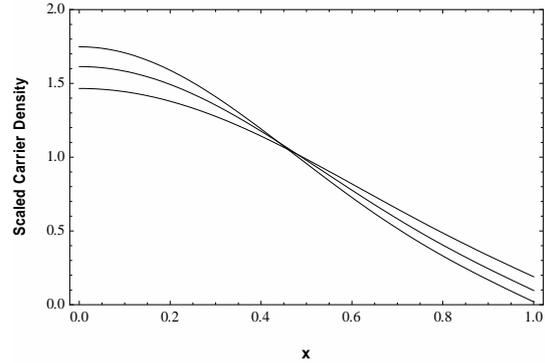

**Figure 5.15**: Carrier density at the end of the pore for β=0.1,0.5,1

**Figure 5.16**: Carrier density over the pore for β=0.1,0.5,1

As observed with the Neumann boundary conditions, the time it takes for the system to empty (i.e. its stationary state equivalent) decreases with increasing $\beta$.

The nonlinear system with Neumann boundary conditions could only be approximated by linear diffusion for T sufficiently close to T=0. With Dirichlet boundary conditions we may approximate the carrier diffusion with a linear system for T sufficiently close to zero, or for T very large. However for the later additional requirements on $\beta$ are needed. To see this, note that for large T the first term in the series for the linear and nonlinear solutions dominates the series and we have therefore

$$\frac{2}{\alpha} v(x,T) = \frac{2\pi^2 \beta \coth(\beta)}{4\beta^2 + \pi^2} \cos\left(\tfrac{1}{2}\pi x\right) e^{-\tfrac{1}{4}\pi^2 T} + O\left(e^{-\tfrac{1}{2}\pi^2 T}\right), \qquad (5.26)$$

and

$$\frac{2}{\alpha} v_L(x,T) = 2\cos\left(\tfrac{1}{2}\pi x\right) e^{-\tfrac{1}{4}\pi^2 T} + O\left(e^{-\tfrac{1}{2}\pi^2 T}\right). \qquad (5.27)$$

In order for the linear system to approximate the nonlinear system we require

$$\pi^2 \beta \coth(\beta) = 4\beta^2 + \pi^2 \qquad (5.28)$$

of which the only solution is β=0. However (5.28) is sufficiently smooth about β=0 and so (5.27) yields a good approximation for large T and β very small. Comparisons of the approximation may be seen in Figures 5.17 to 5.20 which show the nonlinear system (heavy) approaching the linear system (fine) for varying time and β.

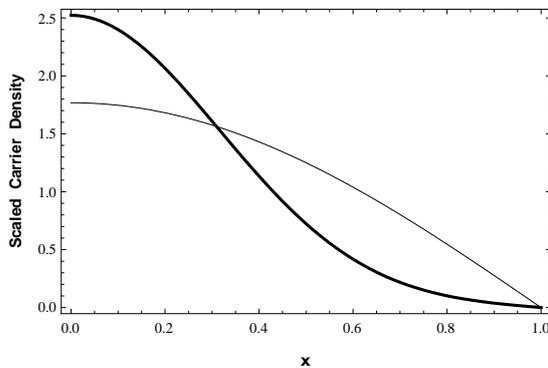 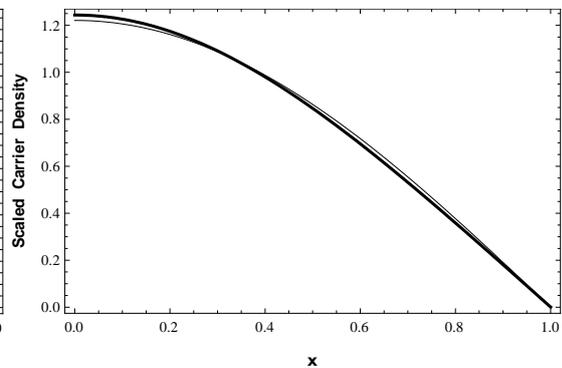

**Figure 5.17**: T=0.05, β=0.002

**Figure 5.18**: T=0.2, β=0.002

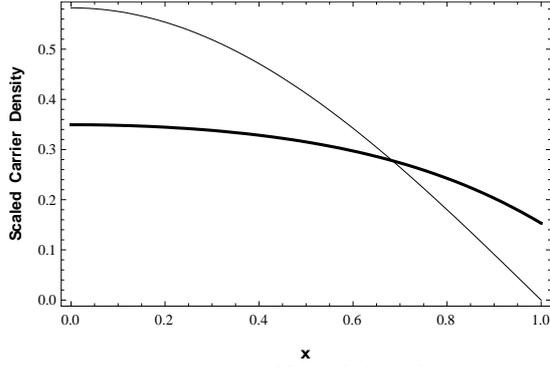 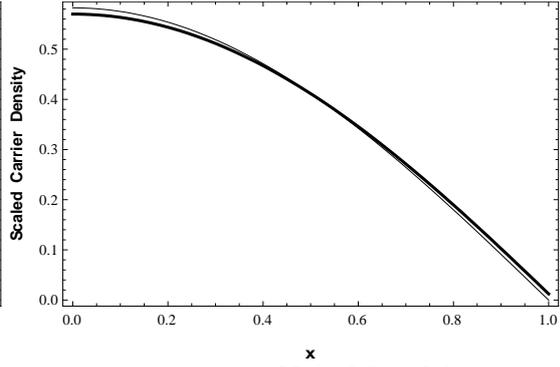

**Figure 5.19**: T=0.5, β=2     **Figure 5.20**: T=0.5, β=0.1

The approximation gains even more accuracy the further we increase T or decrease β to the point where both profiles become quite indistinguishable. From β=α/4γ we note that the nonlinear system approaches linear diffusion for large T when the ratio of initial charge in the system to the diffusive coefficient is small.

Building on what we have illustrated with a specific example we now generalise these observations to an arbitrary symmetric charge distribution satisfying constant Dirichlet boundary conditions in the pore. The scaled carrier density for x over [0,1] is given by

$$\frac{2}{\alpha}v(x,T) = \frac{1}{\beta}\frac{\partial}{\partial x}\left\{\frac{\sum_{k=1}^{\infty}a_k e^{-\frac{1}{4}k^2\pi^2 T}k\pi\sin\left(\frac{1}{2}k\pi x\right)}{a_0 + 2\sum_{k=1}^{\infty}a_k e^{-\frac{1}{4}k^2\pi^2 T}\cos\left(\frac{1}{2}k\pi x\right)}\right\},$$

$$a_k = \int_0^2 \exp\left(\phi_P(x,0)/2\gamma\right)\cos\left(\tfrac{1}{2}k\pi x\right)dx. \qquad (5.29)$$

where $\varphi_P$ is the odd periodic extension of the initial potential. The carrier density for the linear system is

$$\frac{2}{\alpha}v_L(x,T) = \sum_{k=1}^{\infty}a_{L,k}e^{-\frac{1}{4}k^2\pi^2 T}k^2\pi^2\cos\left(\tfrac{1}{2}k\pi x\right),\ a_{L,k} = \frac{2}{\alpha}\int_0^2\phi_P(x,0)\cos\left(\tfrac{1}{2}k\pi x\right)dx \qquad (5.30)$$

where $a_{L,0}$ is zero due to the symmetry of the carrier density. The non-linear system may therefore approximate the linear system for T very large under the general requirement of

$$a_1 - 2\beta a_0 a_{L,1} = 0, \qquad (5.31)$$

and a general formula for g(T) may also be determined as

$$g(T) = -2\gamma\log\left(\frac{1}{2}a_0 + \sum_{k=1}^{\infty}(-1)^k e^{-k^2\pi^2 T}\int_0^2 \exp\left(\phi_P(x,0)/2\gamma\right)\cos(k\pi x)dx\right). \qquad (5.32)$$

We may use this expression to determine g'(T) in the limit of zero diffusivity, i.e. γ→0. We first assume two things; $\varphi_P(x,0)$ has a Taylor expansion over [0,2], about x=0, and that $\varphi_P(0,0)$ is a global maximum which is the case for all examples considered here. Doing so and approximating the potential to second order, noting that $\partial\varphi_P(0,0)/\partial x=0$ due to the symmetry of the problem, one obtains the error function from the integral in (5.32). If the gradient of the potential at x=0 and τ=0 is discontinuous, as is the case for the delta function in the initial

*19*

carrier density, then one can appeal to the Fourier series representation of the extended initial condition to obtain $\partial \varphi_P(0,0)/\partial x = 0$. Using well known asymptotic series expansions for the error function one may then show,

$$\lim_{\gamma \to 0} g'(T) = -\lim_{\gamma \to 0} 2\gamma \frac{1}{\theta_4(0,q)} \frac{d}{dq} \theta_4(0,q), \quad (5.33)$$

where $\theta_4(z,q)$ is the fourth theta function [22], and $q = \exp(-\pi^2 T)$. From the product definition of the theta function we have then

$$\frac{1}{\theta_4(0,0)} \frac{d}{dq} \theta_4(0,0) = -2,$$

which is an upper bound for the strictly decreasing function. Note that this is valid only for $0 \leq q < 1$, which translates to $T > 0$. The $T = 0$ case is obtained by examining the summation under the integral sign in (5.32). Letting $T = 0$ it is may be shown that the summation on the numerator of $g'(T)$ is zero. We have therefore

$$\lim_{\gamma \to 0} g'(T) = 0 \text{ for } T \geq 0, \quad (5.34)$$

which will be useful when examining non-dissipative systems with similar Dirichlet boundary conditions. Note that if $x = 0$ was not the global maximum, the domain of integration could be partitioned in such a way that the maximum in each interval was an end point of the interval, and then a similar analysis applies and the same result follows.

## 6. Exact solution for one component with zero diffusivity

Zero diffusion constants implies what is commonly referred to as the inviscid limit of equation (5.6) which is an inhomogeneous quasi-linear partial differential equation known as Burgers' inviscid equation; namely,

$$\frac{\partial e}{\partial \tau} + e \frac{\partial e}{\partial x} = -f''(\tau), \quad (6.1)$$

where

$$f''(\tau) = -\frac{\partial}{\partial \tau} e(-1,\tau) - e(-1,\tau) \nu(-1,\tau). \quad (6.2)$$

We will uphold a common terminology applied to this equation and refer to the system with zero diffusion coefficients as a non-dissipative system, and the system with non-zero diffusion coefficients as the dissipative system.

There are several reasons why examining solutions in such a limit is interesting. Firstly, for the single component, we may solve (6.1) using the method of characteristics to obtain exact solutions for non-dissipative charge diffusion. Secondly, solutions of the dissipative system may be viewed as those of the non-dissipative system, but with perturbative effects. Knowing how the single component behaves in these limits will aid in understanding the behaviour of multi-component systems in which some or all of the species have zero or near zero diffusion coefficients. We will see, however, that the characteristics of the dissipative and non-dissipative systems can in certain cases be quite different, and we will have to resolve certain ambiguities arising in the solutions.

To begin, equation (6.1) may be solved exactly for a given initial condition by the method of characteristics. Doing so we have

$$x(0) = x(\tau) - e_0(x(0))\tau - \Omega(\tau), \tag{6.3}$$

where

$$e(x,0) = e_0(x) \text{ and } \Omega(\tau) = f(\tau) - f(0) - f'(0)\tau \tag{6.4}$$

The solution of the initial-value-problem of (6.1) may therefore be written as

$$e(x(\tau),\tau) = e_0(x(0)) + \Omega'(\tau), \tag{6.5}$$

where the carrier density is

$$\nu(x,\tau) = \frac{e'_0(x(0))}{(1 + e'_0(x(0))\tau)}. \tag{6.6}$$

As previously observed in [6,7], such density profiles may sustain discontinuities in the density and the field. In keeping with the dissipative systems we will examine only symmetric carrier densities which allows the simplification of (6.3)-(6.5), for in this case $f(\tau)=\Omega(\tau)=0$.

### *6a. Neumann boundary conditions:*

Consider the simple example previously examined in section 5, where now, however, we scale the initial charge within the pore to unity. We have as an initial condition

$$Q_0 = \int_{-1}^{1} \nu(x,\tau)dx = 1,$$

the carrier density in this case being $\delta(x)$. The electric field is therefore governed by

$$\frac{\partial}{\partial \tau}e(x,\tau) + e(x,\tau)\frac{\partial}{\partial x}e(x,\tau) = 0 \quad e(x,0) = \begin{cases} -1/2 & \text{for } x < 0 \\ 1/2 & \text{for } x > 0 \end{cases}. \tag{6.7}$$

Such a discontinuous initial condition in relation to a hyperbolic PDE such as (6.7) forms what is known as a Riemann problem. Consider the characteristics (6.3) for this initial condition and note

$$x(\tau) = \begin{cases} x(0) - \tau/2 & x(0) < 0 \\ x(0) + \tau/2 & x(0) > 0 \end{cases}, \tag{6.8}$$

observing that the characteristics move away from the initial discontinuity, as displayed in Figure 6.1.

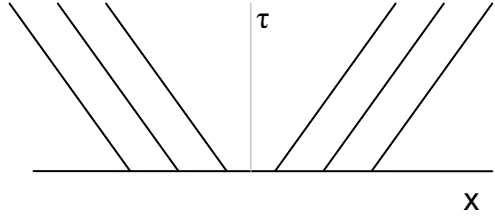

**Figure 6.1**: characteristics for (6.5) showing an ambiguity in the choice of solution at the point of discontinuity

The lack of information for characteristics for x=0, implied by the lack of information within the discontinuity, is an indication that solutions for this type of problem are generally not unique. For example one solution of (6.7) is a stationary discontinuity in the electric field, i.e.

$$e(x,\tau) = \begin{cases} -1/2 \text{ for } x < 0 \\ 1/2 \text{ for } x > 0 \end{cases}. \quad (6.9)$$

However there are indeed other solutions. We must uphold a consistency with solutions of the more general dissipative system and so must therefore consider solutions of (6.7) as the non-dissipative limit ($\gamma \to 0$) of the more general dissipative (5.6). Doing so yields the unique solution under this specific dissipative perturbation [17]. This additional requirement is referred to as an admissibility condition [23]. For the initial condition of a delta function in the charge density, recourse to equations (5.4)-(5.5) imply that the correct solution for the non-dissipative limit is the centered refractive fan, where the electric field takes the form

$$e(x,\tau) = \begin{cases} -1/2 & x \leq -\tau/2 \\ x/\tau & -\tau/2 < x < \tau/2 \\ 1/2 & x \geq \tau/2 \end{cases}. \quad (6.10)$$

To see this, consider pushing the ends of the pore out to infinity. The solution within the pore must then approximate the solution in section 5 which is over the real line. Utilising that solution one may determine the correct profile of the electric field under vanishing dissipation which, while for the entire real line, must also hold as a limiting case.

It may be verified directly that (6.10) satisfies (6.7). It is interesting to note that (6.10) describes a charge density that is a broadening square pulse, and so from the admissibility condition of the non-dissipative limit, we have determined the correct limit representation of the delta function in the initial charge density. Recourse to the non-dissipative limit, while ensuring the correct solution, is not always directly applicable, particularly when considering the more general system of M-species. We therefore appeal to an alternative method for choosing the correct solution, and we do so by observing that (6.7) may be written as

$$\frac{\partial}{\partial \tau}\left\{\frac{1}{2}e^2(x,\tau)\right\} = -e(x,\tau)j(x,\tau), \quad (6.11)$$

where j(x,τ) is the scaled current density. This is the differential form of Poynting's theorem in one-dimension. It is in fact a general statement of energy conservation derivable from Maxwell's equations, where $e^2(x,\tau)/2$ is the energy density of the field and $e(x,\tau)j(x,\tau)$ is the rate of energy imparted from the field to mobile carriers. Charge diffusion is occurring under its own influence which means the energy density of the field within the pore must be a nonincreasing function of time (which may be shown in a purely mathematical setting – see [23] theorem 10.16) and so we have at any time, the total energy of the field being less than or equal to that at a previous time minus whatever has been imparted to the carriers. That is,

$$\int_{x_1}^{x_2} \frac{1}{2} e^2(x,\tau_2) dx \leq \int_{x_1}^{x_2} \frac{1}{2} e^2(x,\tau_1) dx - \int_{x_1}^{x_2} dx \int_{\tau_1}^{\tau_2} e(x,\tau) j(x,\tau) d\tau \text{ for } \tau_2 > \tau_1, \quad (6.12)$$

which is an estimate obtained by integrating (6.11).

This must always hold whether the field is continuous or not, although when the field is discontinuous (6.11) is not applicable while (6.12) generally is. Following [23,24] one may go on to show that solutions obtained via examining the non-dissipative limit of (5.6) satisfy (6.12). Uniqueness theorems from the theory of hyperbolic conservation laws may also be applied. For example, substitution of (6.9) and (6.10) directly into (6.12) shows that only (6.10) satisfies Poynting's theorem and is therefore, due to arguments pertaining to uniqueness (which we won't reproduce here) the solution for the Riemann problem.

We can also determine a rather general result for the Neumann case. Recall that for large $\gamma\tau$, where we fix $\gamma$ and let $\tau\to\infty$, the electric field for a symmetric carrier distribution and constant Neumann boundary conditions is given by (5.14), that is, $2e/\alpha=\omega_0\tan(\omega_0\beta)/\beta$, where $\omega_n\tan(\omega_n)=\beta$ and $\beta=\alpha/4\gamma$. This steady state result is only dependent upon the total amount of charge within the pore. We consider this steady state limit and let $\gamma\to 0$. In this limit $\omega_n = n\pi + \pi/2$ and so

$$e(x,\infty) = \lim_{\gamma\to 0} \pi\gamma \tan\left(\frac{\pi}{2}x\right). \quad (6.13)$$

For any initial symmetric carrier density, the field at the ends of the pore can be made to remain finite while the field within the body of the pore becomes zero. Via an appropriate limiting procedure we can ensure the electric field takes the specified Neumann boundary conditions. However in doing so, the same limit applied to the carrier density results in a singularity within the charge density, existing only at the ends of the pore. To see this, expand (6.13) about x=1, say, and choose $\gamma=(1-x)/4$. Doing so retains the Neumann boundary condition of $e(1,\infty)=1/2$. However using this same prescription for $\gamma$ in the carrier density produces a singularity. We therefore expect that for the non-dissipative Neumann case, the carrier density in the body of the pore empties accumulating at the ends only. This is also the first indication that we need to augment the carrier density with delta functions defined at the ends of the pore to accommodate for this build up of charge.

To illustrate this idea we examine another particularly useful example, that of a unit charge density over the entire pore which has the initial electric field of

$$e(x,0) = \frac{1}{2}x, \quad (6.14)$$

and a solution of (6.7),

$$e(x,\tau) = \frac{x}{(2+\tau)}. \quad (6.15)$$

While there is no ambiguity for this solution, it is however still problematic for it implies that the electric field over the entire pore (including the ends) tends to zero as $\tau\to\infty$ which is contradictory to the boundary conditions. One way to correct this and align the solution with that of the dissipative system in the non-dissipative limit, is to augment the initial carrier density by specifying delta functions at the ends of the pore as previously noted. Consider then,

$$\nu(x,\tau) = \mu(x,\tau) + \Lambda(\tau)\delta(x-1) \text{ for } 0 \leq x \leq 1 \text{ \& } \Lambda(0) = 0, \tag{6.16}$$

where we need only work in the half space due to the symmetry of the system. For Neumann boundary conditions we have at once

$$e(1,\tau) = \int_0^1 \mu(x,\tau)dx + \Lambda(\tau) = \frac{1}{2}Q_0. \tag{6.17}$$

To better address possible discontinuities in the initial carrier density consider a density that is specified over some subset of x belonging to [0,1], say x in [0,$x_c(\tau)$], then from Poisson's equation the electric field may be derived as

$$e(x,\tau) = \begin{cases} \int_0^x \mu(y,\tau)dy & 0 \leq x \leq x_c(\tau) \\ e(x_c(\tau),\tau) + \Lambda(\tau) & x_c(\tau) \leq x \leq 1 \end{cases}. \tag{6.18}$$

Since $e(x,\tau)$ is constant in $x_c(\tau) \leq x \leq 1$, $e(1,\tau) = e(x_c(\tau),\tau)$. However from (6.17) $e(1,\tau) = Q_0/2$ a constant for $x_c(\tau) \leq x \leq 1$. Therefore $\Lambda(\tau) \rightarrow \Gamma(\tau)\theta(\tau - \tau_c)$ where $\theta$ is the Heaviside step function, $\tau_c$ is the critical time when $x(\tau_c) = 1$ and $\Gamma(\tau)$ is yet to be determined. We must also shift the initial condition on $\Gamma$. The carrier density may now be given by

$$\begin{cases} \mu(x,\tau) & 0 \leq x \leq x_c(\tau) \\ \Gamma(\tau)\theta(\tau - \tau_c)\delta(x-1) & x_c(\tau) \leq x \leq 1 \end{cases}, \quad \Gamma(\tau_c) = 0 \text{ and } x_c(\tau) = 1, \tau \geq \tau_c \tag{6.19}$$

where upon substituting (6.16) into (6.6) $\mu(x,\tau)$ is

$$\mu(x,\tau) = \frac{e'_0(x(0))}{(1 + e'_0(x(0))\tau)} \quad |x| \leq x_c(\tau). $$

The weight on the delta function can be determined from observing the behaviour of the current density at the ends of the pore. If we examine a small region about x=1, say, then from the continuity equation (3.8), and the augmented carrier density (6.19), we have

$$j(1-\varepsilon,\tau) = \frac{\partial}{\partial \tau}\left(\int_{1-\varepsilon}^1 \mu(x,\tau)dx\right) + \Gamma'(\tau) \tag{6.20}$$

where $j(1,\tau) = 0$ since, given the boundary conditions, there can be no current flow through the ends of the pore. The continuity of the carrier density $\mu$, where ever it is defined, implies that in the limit $\varepsilon \rightarrow 0$,

$$\Gamma'(\tau) = e(1,\tau)e_x(1,\tau), \tag{6.21}$$

where (3.9) relates the current density to the carrier density and electric field.

Consider again the situation of an initial constant carrier applied over the pore. From equations (6.5),(6.6),(6.21) and the initial condition $\Gamma(0)=0$, the carrier density weight at the end of the pore is

$$\Gamma(\tau) = \frac{1}{2} - \frac{1}{(2+\tau)}, \qquad (6.22)$$

and so by (6.18), noting that in this situation $x_c(\tau)=1$ for all $\tau$, the electric field is given by

$$e(x,\tau) = \begin{cases} \dfrac{x}{(2+\tau)} & 0 \leq x < 1 \\ \dfrac{1}{2} & x \geq 1 \end{cases}. \qquad (6.23)$$

The carrier density by (6.19) is therefore

$$\nu(x,\tau) = \begin{cases} \dfrac{1}{(2+\tau)} & 0 \leq x < 1 \\ \dfrac{1}{2} - \dfrac{1}{(2+\tau)} & x = 1 \end{cases}. \qquad (6.24)$$

We now have a description of the carrier density within the pore that is consistent with the observation (6.13) and which also satisfies the admissibility condition (6.12). That is, it satisfies the non-dissipative limit and (necessarily) the admissibility condition. To see this, for example, note that the current density $j(1,\tau)=0$ and using similar arguments as we did for (6.20), the admissibility condition may be written as

$$\int_{-1}^{1} \frac{1}{2} e^2(x,\tau_2) dx \leq \int_{-1}^{1} \frac{1}{2} e^2(x,\tau_1) dx - \frac{2}{3} \int_{\tau_1}^{\tau_2} e^3(1-\varepsilon,\tau) d\tau$$

for some real $\varepsilon>0$ and arbitrary $x_1$ and $x_2$. Substituting for the electric field and letting $\varepsilon \rightarrow 0$ results in equality.

Observe that (6.23) satisfies the Neumann boundary conditions where the discontinuity in the field is accounted for by the delta function in the carrier density. The dynamics of the diffusion taking place within the pore can be seen immediately; the bulk density remains constant in its distribution, diffusing in a manner as to instantaneously accumulate charge upon the very ends of the pore. The final steady state is therefore a system where half the initial charge is deposited at the ends of the pore and nowhere else. Compare this situation to that observed in section 5 where the inclusion of the diffusion constant yielded steady states where charge concentrated upon the ends of the pore but also remained within the body. The dissipative effects of the second derivative brought about a smooth $\tan(x)$ profile for the electric field in the steady state, where as for $\gamma \rightarrow 0$, an increasing sharpening of the electric field is observed at the ends of the pore eventually becoming discontinuous there in the non-dissipative limit. Physically, this is due to charge deposition at a single point.

We may further extend this example by now considering an initial carrier density that can, via a limit process, take the form of the delta function as well as sustain a discontinuity in the carrier density. We do so by considering an initial square pulse. From (6.10) we know that this is the correct profile under the non-dissipative limit and so we are removing any ambiguity of the weak solution from the outset. Therefore consider,

$$v(x,0) = \frac{1}{\alpha} \text{ for } |x| \leq \frac{\alpha}{2}, v(x,0) = 0 \text{ elsewhere, and } \alpha \leq 2 \qquad (6.25)$$

The initial electric field may be determined as

$$e(x,0) = \begin{cases} \dfrac{x}{\alpha} & 0 \leq x \leq \dfrac{\alpha}{2} \\ \dfrac{1}{2} & \dfrac{\alpha}{2} \leq x \leq 1 \end{cases}. \qquad (6.26)$$

Observe that in the limit $\alpha \to 0$, the electric field becomes (6.7) and the carrier density is a representation of a delta function. We obtain as a solution,

$$e(x,\tau) = \begin{cases} \dfrac{x}{(\alpha + \tau)} & 0 \leq x \leq x_c(\tau) \\ \dfrac{x_c(\tau)}{(\alpha + \tau)} + \Gamma(\tau)\theta(\tau - \tau_c) & x_c(\tau) \leq x \leq 1 \end{cases}. \qquad (6.27)$$

The contribution from the delta function at the ends of the pore is zero at this stage for there is no charge there until $x_c(\tau)=1$. We may determine $x_c(\tau)$ by noting that the discontinuity in (6.27) is caused by a discontinuity in the carrier density. For this particular initial condition the characteristics where charge is defined are parallel to one another and therefore the dynamics of the point of discontinuity within the charge density may be determined for any $\tau$ from the initial condition. Note from (6.3), we have

$$x_c(\tau) = \frac{1}{2}(\alpha + \tau). \qquad (6.28)$$

The discontinuity in the carrier density therefore advances towards the ends of the pore reaching the ends at the critical time

$$\tau_c = 2 - \alpha. \qquad (6.29)$$

At this time the step function in (6.19) becomes unity and the weight upon the delta function, augmenting the carrier density at the ends of the pore, comes into play. We find this weight using (6.21) and the initial condition $\Gamma(\tau_c)=0$; namely

$$\Gamma(\tau) = \frac{1}{2} - \frac{1}{(\alpha + \tau)}, \quad \tau \geq \tau_c. \qquad (6.30)$$

We now have a complete description of the carrier diffusion in the pore for the initial square pulse. Note that the initial carrier distribution diffuses over the pore, maintaining a discontinuity in the carrier density as shown diagrammatically in Figure 6.2, until the system reaches the critical time (6.29).

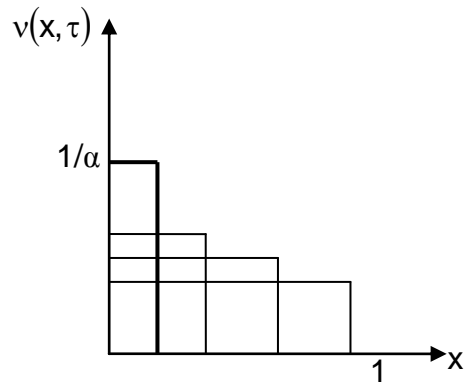

**Figure 6.2**: Diagrammatic representation of square pulse diffusing over half the pore

At that critical time there is then a charge density at the ends of the pore and the system is in a situation previously considered where a constant carrier density is over the entire pore. We therefore expect the same description for the electric field and carrier density as that given in (6.23) and (6.24). As may be shown quite easily, this is indeed the case.

Note that as $\tau \to \infty$ the field within the pore becomes zero and all charge is deposited upon the ends, satisfying the general observation of (6.13). One may show that the electric field (6.27) satisfies the estimate of Poynting's theorem (6.12) and is therefore the physically relevant solution.

### *6a. Dirichlet boundary conditions:*

We conclude this section with a brief discussion of the same initial conditions where the potential at the ends of the pore is maintained at zero. The dynamics of these initial charge profiles under Dirichlet conditions are essentially the same as those of the Neumann case except the system must empty completely. The ends of the pore therefore require no specification of a delta function (consider that the ends of the pore are grounded and so no charge accumulation may occur), as the field at the end of the pore is zero. We also know this because it was shown that g'(T)=0 for all T (5.34). As was the case for the dissipative system we must therefore create an odd periodic extension of the initial charge density in order for the boundary conditions to be satisfied. Ambiguities in choosing the correct weak solution of the non-dissipative system may be resolved by considering Poynting's theorem over x in [-1,1] which may then be extended over the real line.

Consider then the initial condition (6.7). The extended initial electric field is shown in Figure 6.3.

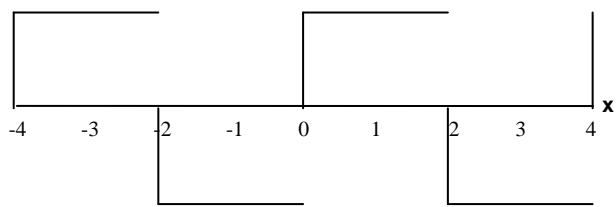

**Figure 6.3:** Periodic array of image charges (vertical spikes), and electric field, establishing boundary conditions on the physical charge in x ε [-1,1].

The admissibility condition requires that the electric field for x over [-1,1] be a refractive wave taking a form similar to that of (6.10), and so the charge density evolves as a periodic set of broadening square pulses, emanating from the initial points of discontinuity of the electric field. When the charge density reaches the ends of the pore it is met with a negative image charge density. This ensures that no accumulation of charge can occur at the points x=1+2n, n=±1,±2,…. At this time there is a constant charge density over the entire pore, and the periodic extension is a square wave as illustrated in Figure 6.4. The density then diffuses in a similar fashion to density described by (6.15), which is simply the reduction of the amplitude of the square wave. As $\tau \to \infty$ it may be seen, by considering the dissipative system, that the amplitude of the wave becomes zero and so the pore empties completely.

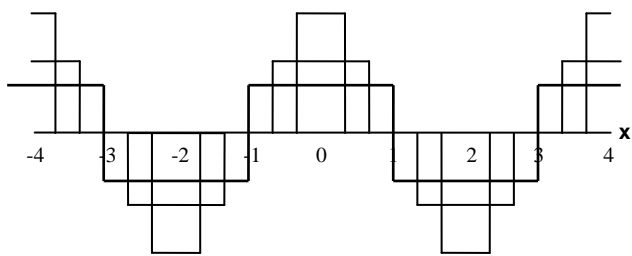

**Figure 6.4:** Depiction of the periodic extension of the charge density diffusing to a square wave.

## 7. Discussion

Beginning with the physically applicable assumption of a cylindrical pore with a dielectric constant much greater than its surrounds, we have shown that the cumulative electric potential of M-species within the pore obeys the one-dimensional Poisson equation. Upon inferring a Nernst-Planck relation for the species within the pore we are then lead to a set of coupled nonlinear partial differential equations that were somewhat simplified by scaling. Up to this point we essentially reproduced the results of Barcilon et al., although here we assumed induced charge on the surface of the pore produced a field that was negligible compared to that produced by the carriers within the pore. Several interesting and general observations were then made concerning the system and the different boundary conditions of Neumann and Dirichlet. In particular it was observed that for Neumann boundary conditions, not all species could obtain a constant density when in a steady state. It was shown for the Neumann case that comparisons with the dissipative linear model of diffusion could only be made near the initial condition, and that the long term behaviour of the non-linear system was markedly different. Whereas for the Dirichlet case, diffusion in the dissipative system approximated that of the linear system for times both small and large, although for the latter additional requirements are needed on the diffusion coefficient. It was observed that for the symmetric boundary values examined here, all charge leaked out of the pore for the Dirichlet boundary condition while for the Neumann boundary condition, all charge remained confined to the pore.

The same initial conditions examined in the dissipative case yielded ambiguities in the non-dissipative system. To obtain the physically relevant solution, recourse was made to the dissipative solution in the non-dissipative limit and discontinuities in the initial charge density were seen to propagate due to a rarefactive fan in the electric field. The method of determining the appropriate solution via vanishing viscosity was somewhat generalised utilising an estimate from Poynting's theorem and results from the theory of hyperbolic equations. A further interesting observation is obtained by (following [23,24]) re-writing Poynting's estimate (6.11)-(6.12) as

$$\frac{\partial}{\partial \tau}\left\{\frac{1}{2}e^2(x,\tau)\right\} + \frac{\partial}{\partial x}\left\{\frac{1}{3}e^3(x,\tau)\right\} \leq 0, \tag{7.1}$$

which then defines a so called entropy/entropy-flux pair, which for our purposes here may be interpreted as

$$\int_{x_1}^{x_2}\frac{1}{2}e^2(x,\tau_2)dx \leq \int_{x_1}^{x_2}\frac{1}{2}e^2(x,\tau_1)dx - \frac{1}{3}\int_{\tau_1}^{\tau_2}e^3(x_2,\tau)d\tau + \frac{1}{3}\int_{\tau_1}^{\tau_2}e^3(x_1,\tau)d\tau \tag{7.2}$$

It may be observed that any solution satisfying (6.7) that has a discontinuity must also satisfy the Rankine-Hugoniot condition. This condition when applied to the entropy/entropy-flux pair estimate yields a contradiction as may be seen in LeFloch [24, pg.12]. This in turn implies that solutions satisfying Poynting's theorem must have a continuous electric field. However we have already shown via the non-dissipative limit, steady state solutions are those in which all carriers exist at the ends of the pore and nowhere else. This produces a steady state discontinuous electric field, and hence an apparent contradiction. The problem lies with (7.1) and the fact that it doesn't explicitly take into account a zero current density at the ends of the pore. If one does account for the current density, then any discontinuity within the field when used in the estimate is restricted to a single point at the end of the interval of integration. This makes no contribution to the integral and so no contribution to the estimate. The entropy/entropy-flux pair estimate of (7.2) fails to do this in an explicit fashion. Instead, a discontinuity in the electric field within the estimate persists beyond the ends of the pore (due to the boundary condition) and so makes a contribution which then violates Poynting's theorem. However (7.2) and the afore mentioned consequence of the electric field necessarily being continuous still applies, though only for regions where the current density is non zero. For in such regions (7.1) may be derived directly from (6.12).

While this paper examines M-component charge transport in one dimension, applicable to a nanopore, or indeed many general electrodiffusive situations in one dimension (via averaging), it has focused primarily on the dynamic solutions for a single component, specifically the case of symmetric boundary conditions. We have therefore preliminary results for comparison against the more interesting system dealing with M-components, particularly a non-dissipative system perturbed to then incorporate the diffusive term within the Nernst-Planck equations.

## *Acknowledgement*

We thank John Liesegang for discussions regarding this work as well as the valuable and constructive comments from the anonymous reviewers. Writing now as the first author, I would like to acknowledge and thank the second author, Edgar R Smith, who sadly passed away in 2009. Ed was a brilliant mathematician and teacher and a very good friend to many people. He was an extremely generous person and is sorely missed. Ed had a deep passion for exploring the ideas and theories driving physical phenomena, and while he was primarily concerned with the complicated nature of colloids and other statistical mechanics, he was never shy in rolling up his sleeves to work on more rudimentary problems like the one presented here. He took great pleasure in this and I think he would have been rather pleased with the final results in this paper. Cheers Ed.